\documentclass{aa}  

\usepackage{graphicx}
\usepackage[breaklinks,colorlinks,urlcolor=blue,citecolor=blue,linkcolor=blue]{hyperref}
\usepackage{multirow}
\newcommand{\MYhref}[3][blue]{\href{#2}{\color{#1}{#3}}}
\usepackage{txfonts}

\newcommand{\nuobs}{$\mathrm{\nu_{obs}}$}

\newcommand{\Rs}{R$_*$}

\newcommand{\nup}{$\mathrm{\nu_p}$}
\newcommand{\nuB}{$\mathrm{\nu_B}$}
\newcommand{\nupr}{$\mathrm{\nu_p(r)}$}
\newcommand{\nuBr}{$\mathrm{\nu_B(r)}$}
\newcommand{\TB}{$\mathrm{T_B}$}

\usepackage{colortbl}
\definecolor{Grey}{rgb}{0.5,0.5,0.5} 
\definecolor{MyRed}{rgb}{0.9,0.0,0.0} 
\definecolor{MyPink}{rgb}{0.8,0.3,0.5} 
\definecolor{MyMediumBlue}{rgb}{0.7,0.72,1.0} 
\definecolor{MyGreen}{rgb}{0.0,0.9,0.5} 
\definecolor{SWGreen}{rgb}{0.0,0.5,0.0} 
\definecolor{SWRed}{rgb}{0.9,0.0,0.0} 
\definecolor{AMpurple}{rgb}{0.9,0,1.0} 
\definecolor{AMbrown}{rgb}{0.8,0.3,0.3}

\begin{document}

    \title{Energetic particle activity in AD Leo: Detection of a solar-like type-IV burst} 
   \subtitle{}
    \titlerunning{Detection of a type-IV radio burst in AD Leo}
    
   \author{Atul Mohan\inst{1,2,3,4}
         , Surajit Mondal\inst{5}
         , Sven Wedemeyer\inst{1,2}
        , Natchimuthuk Gopalswamy\inst{4}        }
        \institute{Rosseland Centre for Solar Physics, University of Oslo, Postboks 1029 Blindern, N-0315 Oslo, Norway
    \and
    Institute of Theoretical Astrophysics, University of Oslo, Postboks 1029 Blindern, N-0315 Oslo, Norway
    \and
    The Catholic University of America, 620 Michigan Avenue, N.E. Washington, DC 20064, USA
    \and
    NASA Goddard Space Flight Center,  8800 Greenbelt Road, Greenbelt, MD, 20771, USA
    \and
    Center for Solar-Terrestrial Research, New Jersey Institute of Technology, 323 M L King Jr Boulevard, Newark, NJ 07102-1982, USA\\
    \email{atul.mohan@nasa.gov}
    }

             
             
\authorrunning{Mohan, A. et al.}

   \date{Received September 15, 1996; accepted March 16, 1997}
   \date{Received  - ; accepted  - }

 
  \abstract
  {
   AD Leo is a young and active M dwarf with high flaring rates across the X-ray-to-radio bands. 
   Flares accelerate particles in the outer coronal layers and often impact exo-space weather. Wide-band radio dynamic spectra let us explore the evolution of particle acceleration activity across the corona. Identifying the emission features and modelling the mechanisms can provide insights into  the possible physical scenarios driving the particle acceleration processes.  
   }
 {We performed an 8\,h monitoring of AD\,Leo across the 550 - 850\,MHz band using upgraded-Giant Metrewave Radio Telescope (uGMRT). The possible flare and post-flare emission mechanisms are explored based on {the evolution of flux density and polarisation.}}
    {The python-based module, Visibility Averaged Dynamic spectrum (VISAD), was developed to obtain the visibility-averaged wide-band dynamic spectra. Direct imaging was also performed with different frequency-time averaging. Based on existing observational results on AD Leo and on solar active region models, radial profiles of electron density and magnetic fields were derived. Applying these models, we explored the possible emission mechanisms and magnetic field profile of the flaring active region.}
  {
The star displayed a high brightness temperature (\TB $\approx$\,10$^{10}$\,--\,10$^{11}$\,K) throughout the observation. The emission was also nearly 100\% left circularly polarised during bursts. The post-flare phase was characterised by a highly polarised (60 - 80\%) {solar-like type IV burst} confined above 700\,MHz.}
   {The flare emission favours a Z -mode or a higher harmonic X-mode electron cyclotron maser emission mechanism. The $>$ 700\,MHz post-flare activity is consistent with a type-IV radio burst from flare-accelerated particles trapped in magnetic loops, which could be a coronal mass ejection (CME) signature. This is the first solar-like type-IV burst reported on a young active M dwarf belonging to a different age - related activity population (`C' branch) compared to the Sun (`I' branch). {We also find that a multipole expansion model of the active region magnetic field better accounts for the observed radio emission than a solar-like active region profile.}
   }


   \maketitle
%

\section{Introduction}
Low-mass main-sequence stars of spectral type M (M~dwarfs) make up the most numerous stellar type in our galaxy. {They also have the highest occurrence rate} of Earth-like exoplanets in their habitable zones~\citep{Bashi20_occurrence_of_smallplanetsFGK}. 
Meanwhile, M-dwarfs form the most active population among {cool main-sequence stars with relatively high values of flaring rates}, X-ray-to-bolometric flux ratio, and other activity metrics (see \citealp{Donati09_Rev_Bfield} for a review).
Moreover, the younger stars are more active with stronger magnetic fields and faster rotation rates~\citep[e.g.][]{Vidotto14_B_Vs_age_n_rot,Lund20_stelar_helicity}.
 Stellar flares cause energetic space-weather events including the release of energetic particles, intense  high-energy radiation and large-scale eruptions, which threaten the stability of planetary atmospheres and hence the habitability. So, it is important to understand the mechanism, evolution, and space-weather consequences of M~dwarf flares.

For this study, we chose the very active M3.5V star AD Leo located at $\approx$ 5 pc from the Sun. The star is proposed to host a planet~\citep{2018AJ....155..192T}, though the detection was later refuted~\citep{2020A&A...638A...5C}. 
The star has a typical mean optical flaring rate of $\approx$ 0.6 h$^{-1}$~\citep{dal2020flare}. 
AD Leo displays high levels of chromospheric and coronal activity in several spectral lines and X-ray bands~\citep{2006A&A...452..987C}. \citet{2019ApJ...871..214V} estimated a radio flaring rate of $\approx$ 0.2 -- 0.3 h$^{-1}$ with a coherent burst emission duty cycle of $\approx$ 20 -- 25\% in the 0.34 -- 1.4\,GHz range. 

{Flare emission in the metrewave band is sensitive to various coherent emission signatures from flare-accelerated electrons either trapped in low-coronal magnetic field structures or streaming across coronal layers~\citep[e.g.][]{wild1970,McLean_solar_radiobook,Bastian90_FlarestarsVLA,Melrose09_CoherentEmiss,melrose2017, Zic20_typeIV_ProximaCen, 2021MNRAS.500.3898V}.}
The two prominent coherent emission mechanisms discussed in the literature include the electron cyclotron maser emission (ECME) and the plasma emission~\citep[][]{Dulk85_FlareradioEmissMech,guedel02_Rev_stellarRadioEmiss,melrose2017}.
ECME dominates when the ratio of plasma frequency to gyro-frequency (\nup/\nuB) at the flare source region is less than 1.5~\citep[]{Melrose82_ECMEinsun_n_stars,Melrose84_ECMEgrowthrate_damp,Dulk85_FlareradioEmissMech,dulk87_ECME_evidance_sun_n_stars}.
When \nup/\nuB$>$3, the favoured emission mechanism is plasma emission~\cite[e.g.][]{ginzburg1958,Tsytovich69,melrose1972,cairns85_secondHarmplasmaemiss,Melrose09_CoherentEmiss}.
Depending on the emission mechanism, the coherent emission frequencies correspond to either the fundamental mode or the harmonics of the local plasma frequency (\nup) or gyro-frequency (\nuB), letting us infer either the local density or magnetic field strength, respectively. 
Apart from the flare emission, M dwarfs are also known to display non-flaring (or non-impulsive) continuum emission with a high brightness temperature (\TB)~\cite[e.g.][]{White89_VLA1GHz_starcatalog,Bastian90_FlarestarsVLA}. In a few GHz (<2\,GHz) bands, the continuum \TB\ ranges from $\approx$10$^7$ - 10$^9$\,K~\citep{Osten2008,2019ApJ...871..214V}. In metrewavebands, \TB\ can be even higher~\cite[e.g.][]{Bastian94_stellarflaresRev,guedel02_Rev_stellarRadioEmiss,2019ApJ...871..214V}.
The continuum emission {can form either from a gyrosynchrotron mechanism,} or a combination of various coherent emission processes operating at multiple active regions \citep{Bastian96_radioflares-sol-stelConn,Benz96_Conerentflares_decimeter,Zaitsev97_decim_continua_Coherentmech}.

There have been many studies that explored radio emission from AD Leo, reporting several flaring events. While instrumental limitations forced earlier studies to use only a few {discontinuous} narrow frequency bands~\citep[e.g.][]{1974ApJ...194L..43S,Bastian90_ADLeoflare_arecebo},
recent observations with continuous wide-band instruments have enabled spectroscopic explorations revealing rich variability in the dynamic spectrum~\cite[e.g.][]{Osten2008,2019ApJ...871..214V,Jiale2023_FAST_ADLeoflares}.
These studies have revealed several features such as fast pulsations at scales of $\approx$\,ms - min and high drift rates in the dynamic spectrum, often related to the propagation of energetic particles or spatio-temporal evolution of plasma instabilities in the associated active regions~\citep[e.g.][]{Bastian90_ADLeoflare_arecebo,Osten2006,Osten2008,Jiale2023_FAST_ADLeoflares}.
However, much of these observations had been close to or above 1 GHz.
Here, we present an 8\,h monitoring of AD Leo across a wide sub-Giga Hertz band, which is expected to probe higher atmospheric layers than those probed by Giga Hertz bands. 
Though the original study was planned with simultaneous observation spanning from the X-ray to the radio band, only the radio observations were successful due to various technical issues. We present the results from metrewave observations here.


Section ~\ref{sec:method} provides the data and analysis methodology. Section~\ref{sec:results} presents the different activity periods observed and the emission characteristics. Section~\ref{sec:discussion} provides our inferences, and Sect.~\ref{sec:conclusion} summarises the conclusions.

\section{Data and methodology} 
\label{sec:method}

AD\,Leo was observed with uGMRT~\citep{2017CSci..113..707G} in Band~4 (550 - 850\,MHz) from 3 December 2021, 21:17:41 UT to 4 December 2021, 04:17:41 UT at $\approx$5\,s and $\approx$100\,kHz time and frequency resolutions, respectively. The data were gathered in RR and LL polarisations, letting us explore the spectro-temporal evolution of the circular polarisation along with brightness temperature (\TB) or flux density. The data analysis was done using the functionalities in the interferometric analysis package, Common Astronomy Software Applications~\citep[CASA;][]{CASA2022}. The data were first flagged and calibrated using standard procedures. 
{The flux calibrator was 3C147, and the phase calibrator was 1021+219.
Portions of data affected by radio frequency interference from terrestrial sources and systematic instrumental effects were identified and flagged manually. The automatic flagging routine in CASA, namely \texttt{tfcrop}, was also used. This was followed by bandpass calibration using the flux calibrator, and then by time-dependent gain calibration using the phase calibrator. The bandpass solution and the time-dependent gain solution were then extrapolated to the target star field. The calibrated target field data were further flagged using the \texttt{Rflag} routine in CASA.}
Initial imaging was done using the CASA task \texttt{tclean} in \texttt{mtmfs} mode, with \texttt{nterms} set to two and averaging over the entire observing time. {This produced a mean model image for the entire spectral band, which}
 was taken as the initial model, and self-calibration was performed in an iterative manner~\citep{cornwell99} to improve it. 
After the second iteration, an optimal model was obtained.
So, we stopped at the second iteration and adopted the CLEAN model of the iteration as the mean model image of the field of view for the observation period.
The AD Leo emission at the phase centre was masked from the final model image, and the model visibilities were subtracted from the corrected data. 
The remaining corrected visibilities pertain to purely stellar emission.
This stellar data were subjected to two forms of analysis as described below. 

First, we developed a python-based package, Visibility Averaged Dynamic spectrum (VISAD), to extract the visibilities from a calibrated CASA measurement set (MS) and generate a dynamic spectrum (DS) with required time and frequency averaging. The package has routines to perform additional data flagging on the calibrated MS if required. The package uses CASA tasks inside a python environment and employs the parallel processing tools in python (multiprocessing\footnote{\MYhref{https://docs.python.org/3/library/multiprocessing.html\#reference}{https://docs.python.org/3/library/multiprocessing.html\#reference}}) enabling an analysis of several hours of wide-bandwidth datasets within a couple of minutes depending on the available machine cores. The module is made publicly available via GitHub\footnote{\MYhref{https://github.com/atul3790/Visibility-averaged-DS/releases/tag/VisAD_V0}{https://github.com/atul3790/Visibility-averaged-DS}}.
We obtained both dynamic spectra and band-averaged light curves in Stokes I and V using VISAD.
{The Stokes V data were corrected as per the suggestions in ~\cite{Barnali20_GMRT_StokesV_conv} to adhere to the standard IAU/IEEE convention.}

We also produced image spectral cubes for the entire observing period with varying time and frequency averaging, which were carefully chosen to detect the star with a signal-to-noise ratio (S/N) of at least five in every spectro-temporal slice. This exercise generated two data sets: a library of stellar spectra for different time bins spanning the entire observation period and a band-averaged stellar light curve.
The spectra were generated with a resolution of 20\,MHz and a time averaging of $\approx$ 2 - 3\,min during the very active periods.
During periods of weak activity, a 50\,MHz averaging with 5 - 15\,min time averaging was employed.
We used the image cube information to cross-verify the flux densities obtained from a visibility-averaged DS for the flare period. The values matched within $\approx$15 - 20\%.
{A source of this discrepancy is the additional flagging done in a couple of higher frequency channels while using VISAD}. The presence of artefacts away from the source would also play a role in the observed difference in the flux density.
In this work, we discuss the spectro-temporal nature of the observed activity, {the possible} physical mechanism, and infer physical properties of the emission region in the corona.

\section{Results}
\label{sec:results}

\subsection{Results from VISAD} \label{sec:visad}
\begin{figure*}
\centering
\includegraphics[width=0.8\textwidth,height=0.7\textheight]{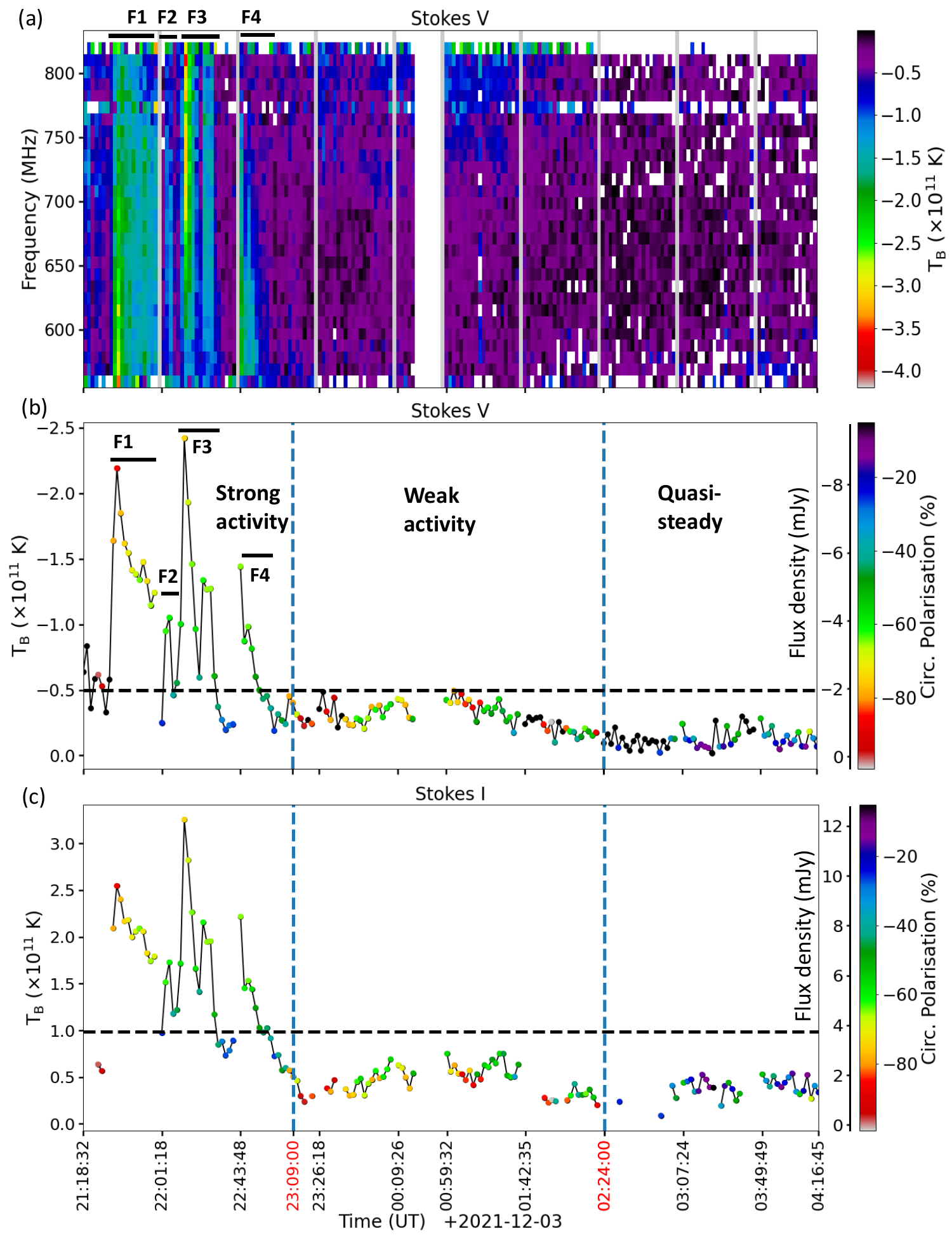}
\caption{Results from VISAD pipeline. (a): Stokes V dynamic spectrum. (b-c): Band-averaged light curves in Stokes V and I. {The 
ordinate 
shows T$_\mathrm{B}$ (left) and flux density (right), respectively. Different periods of activity are marked. Basal flux density levels in Stokes I and V are represented by dotted lines in the respective panels. We note that the $\mathrm{T_B}$ values are nominal and simply highlight that the emission is via a coherent emission mechanism. Since the true source sizes are unknown,  an average (i.e. across the whole unresolved stellar disc)  $\mathrm{T_B}$ is computed here.}}
\label{fig1:DS_LC}
\end{figure*}
\label{sec:results}
\begin{figure}
\centering  
\includegraphics[width=0.5\textwidth]{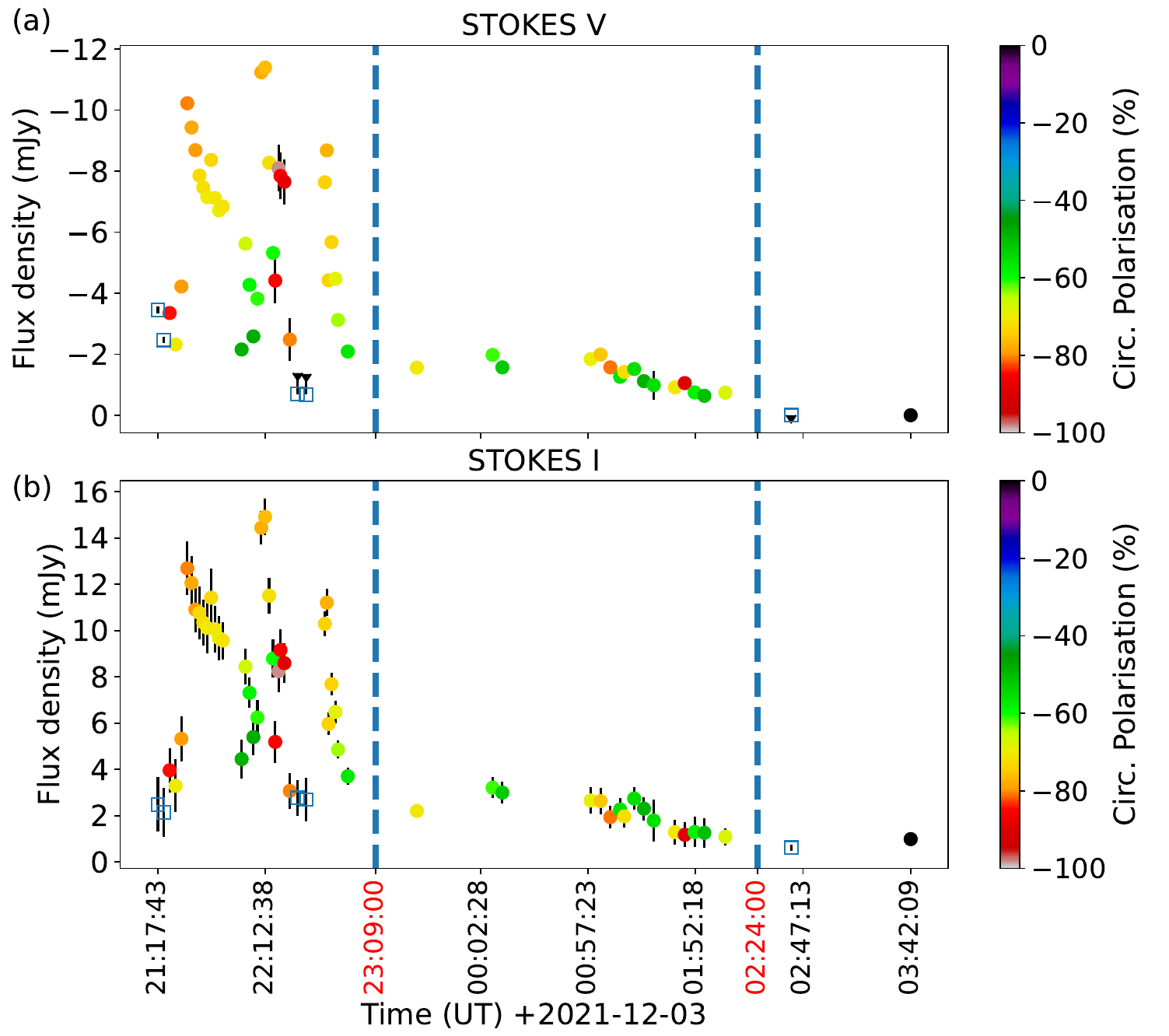}
\caption{Band-averaged Stokes I and V light curves obtained from imaging analysis. The detections below 5$\sigma$ are marked as upper limits. Transparent squares mark poor polarisation estimates.}
\label{fig2:LC_imging}
\end{figure}
Figure~\ref{fig1:DS_LC}a shows the Stokes V dynamic spectrum from VISAD made with 100\,s and 10\,MHz averaging in the top panel. The bottom panels shows the band-averaged light curve obtained from VISAD in Stokes I and V, with 100\,s averaging. 
 Until 23:09:00 UT, we find multiple epochs of strong intermittent flaring activity, during which the stellar {\TB\ suddenly rises }beyond 
10$^{11}$\,K with circular polarisation levels between  60 and 90\%. 
We note that the \TB\ values reported are averaged over the entire stellar disc, thus the true \TB\ will be higher depending on how small the area  of the active region is compared to the size of the star.
The {flare rise phases are sudden and are followed} by decay phases in both \TB\ and polarisation fraction, as clearly seen in the band-averaged light curves.
We mark this initial 2\,h period as a period of strong activity.
Analysing the typical flux levels to which Stokes V emission falls back after the sudden impulsive rise in the light curve, a basal Stokes V flux level is fixed at -2\,mJy, {corresponding to \TB\ =-0.5$\times10^{11}$\,K.} 
A similar exercise on the {Stokes I light curve leads to a basal flux level of $\approx$ 4\,mJy (\TB\ =1.0$\times 10^{11}$\,K),} which is about twice the typical quiescent flux seen during the quasi-steady period, as noted in Fig. \ref{fig1:DS_LC}.
The basal flux levels are used to define flaring epochs during the strong activity period. A flaring epoch is said to begin when the stellar flux rises above the basal flux level and ends when the flux falls back below the basal level.
Based on this definition, four flaring epochs (F1 - 4; see  Fig.~\ref{fig1:DS_LC}b-c) with flare decay timescales of $\approx$ 5 - 30\,min are identified in both of the Stokes light curves.
\begin{figure}
\centering  
\includegraphics[width=0.5\textwidth]{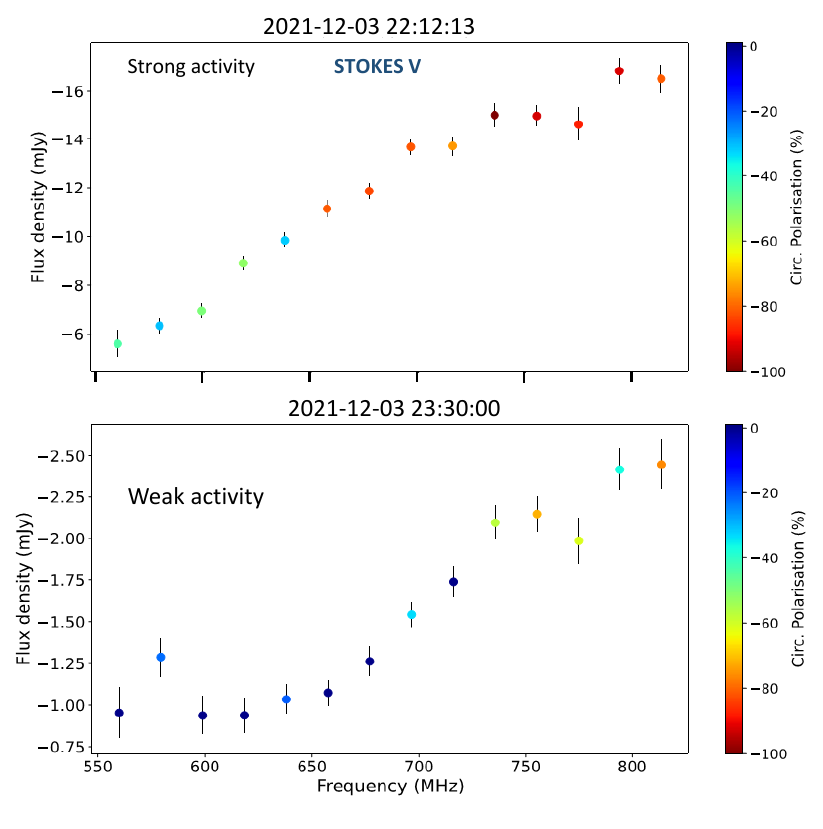}
\caption{Stokes V spectrum from imaging. (a): A 2 min averaged spectrum corresponding to the strongest flare F3. (b) A 51 min averaged spectrum during the for the weak activity period.}
\label{fig3:V_spectra}
\end{figure}


After the strong activity period, the star goes through $\approx$ 1.4\,h period of weak activity between 2021-12-03 23:09:00 UT and 2021-12-04 00:24:00 UT (see Figs.~\ref{fig1:DS_LC}b-c).
This period is characterised by high Stokes V \TB$\approx$0.5 - 1.5$\times$10$^{11}$, primarily the in higher frequency bands ($\gtrsim$700\,MHz) as seen in the VISAD spectrum (Fig.~\ref{fig1:DS_LC}a). 
The observed Stokes V and Stokes I \TB\ is much brighter than the expected coronal temperature of $\approx$3\,MK based on X-ray {studies}~\citep[e.g.][]{vandenBasselaar03_ADleoXrayQF, Namekata20_ADLeoXrayFlares}. Additionally, the band-averaged flux light curves in both Stokes I and V demonstrate high polarisation levels of $\approx$50 - 90\% during the weak activity period, with a gradually varying profile resultant from the dynamic spectral structure of the emission. Such emission characteristics after flares have been observed in the Sun as well and are interpreted as signatures of high-energy particles trapped in post-flare loops~\citep{Gergely86_typeIV_CME_rel_rev, Gopal11_PREconf, Hillaris16_typeIV, Morosan19_typeIV_emissmech, Salas20_typeIVPoln_struct}. We believe that the same phenomenon is also behind the observed emission characteristics in this case.


After the weak activity period, {we find that the Stokes V and I light curves tend to fluctuate around characteristic mean flux values of $\approx$-0.5 and 1.5\,mJy with relatively low circular polarisation levels.} We define this phase as the quasi-steady phase.

\subsection{Results from imaging analysis}\label{sec:imaging}
To independently verify the flux density and polarisation trends presented in Section \ref{sec:visad}, we analysed the band-averaged light curve and image spectral cubes obtained from direct imaging (described in Section \ref{sec:method}). 
Figure~\ref{fig2:LC_imging} shows the imaging-based light curve in Stokes I and V.
Stokes V images have better S/N than those in Stokes I as evident from the lower error bars, {due to relatively low levels of artefacts. Additionally, the number of galactic and extra-galactic sources with a circular polarisation fraction exceeding $10\%$ are significantly low~\citep[e.g.][]{saikia88_extragal_circpol,Macquart00_scintill_induced_circpol,Enslin17_galactic_circpol,callingham23_vLOTSS}.}
Apart from the few time steps where only 5$\sigma$ upper limits could be determined either in V or I maps, we have good polarisation estimates across the light curve. 
As mentioned in Sect.~\ref{sec:method}, overall the fluxes from VISAD and imaging procedures agree within 20\%. 
This flux uncertainty does not affect our analysis or physical modelling in Sect.~\ref{sec:discussion}. This is because, the emission being highly polarised in the flare and weak activity phase, is likely via some coherent emission processes. {These coherent emission processes} are often subject to a host of intrinsic physical processes, which are hard to nail down with just the available radio data (see, \citealp{Bastian90_FlarestarsVLA,melrose2017_coherentEmissMech} and \citealp{2021MNRAS.500.3898V} for an overview). 
Our analysis in Sect.~\ref{sec:discussion} thus requires absolute flux estimation accuracy only within an order of magnitude in \TB~\citep[][]{Bastian90_ADLeoflare_arecebo}.
The strong activity light curve matches well with the VISAD results, and the significantly polarised weak activity period that followed is also well captured in the image-based light curve.
During the weak activity period, Stokes I flux peaks at $\approx$ 2.5\,mJy with a polarisation of $\approx$60\% similar to that seen in the VISAD light curve (Fig.~\ref{fig1:DS_LC}(c)). 
A gradual decline phase after $\approx$00:40:00 UT is also seen in the Stokes I and V light curves  (Fig.~\ref{fig2:LC_imging}a-b), akin to the VISAD light curves (Fig.~\ref{fig1:DS_LC}(b-c)).

In order to understand the spectral evolution of stellar emission, the stellar spectra derived from image spectral cubes described in Sect.~\ref{sec:method} were analysed. Figure~\ref{fig3:V_spectra} shows example Stokes V spectra from specific time steps during the strong and weak activity period. Polarisation fraction is colour-coded.
The spectrum shown in Fig.~\ref{fig3:V_spectra} for the strong activity period is chosen to coincide with the peak epoch of the strongest flare (F3).
The Stokes V flux of the star rises to around 16\,mJy ($\approx4\,\times\,10^{11}$\,K) at high frequencies during this timestep (2021-02-03 22:12:13 UT) as seen in VISAD dynamic spectrum. Circular polarisation reaches 100\% during this time.
The spectrum during the weak activity period shows an enhanced brightening with a significantly enhanced circular polarisation in the high-frequency bands ($\gtrsim$700\,MHz). This agrees with the observed enhanced \TB\ confined to high-frequency ranges in the VISAD dynamic spectrum during the weak activity phase (see Fig.~\ref{fig1:DS_LC}).

\section{Discussion}\label{sec:discussion}
The high \TB\ and {polarisation fraction seen} in the strong activity period suggest a coherent emission mechanism is at play, either via an ECME mechanism or a plasma emission process (see, \citealp{Melrose93_ECMEnplasma_highTB_polflares} for an overview). The following subsections summarise our current understanding of the two mechanisms.
\subsection{Plasma emission}
A plasma emission mechanism is the other coherent emission process, which can also generate highly polarised emission~\citep{Tsytovich69,melrose1972}. However, this mechanism is often subject to near-source scattering~\citep[e.g.][]{steinberg1971,robinson_scat1983,Arzner1999,Atul19_typeIIIQPP_turb,Kontar19_Arznercopy,Chen20_snapshottypeIIIEvol_turb}, which reduces the observed polarisation levels. In the case of solar type-III bursts, which are well-known plasma-emission sources, the maximum Stokes V/I detected is $\approx$ 60\%~\citep{Hilaire,Rahman20_typeIIIstats}. 
However, these results are at frequencies below $\approx$ 100 MHz. \cite{Atul21_typIII_vs_ht} showed that the effects of scattering could be less significant at higher frequencies since the effective scattering screen widths decrease towards higher frequency or equivalently at lower heights above the corona. {This means that the scattering-induced depolarisation at our observing frequencies could be lower than that observed close to $\approx$ 100\,MHz, but its effects may not be negligible. It is therefore unlikely that we observe radio bursts with polarisation levels above $\approx$75\% generated purely via a coherent plasma emission mechanism.}

\subsection{Electron cyclotron maser emission }
\label{sec:ECME}
{An ECME is driven by a loss-cone distribution of electrons. Such an unstable electron pitch-angle distribution can form in post-flare coronal loops~\citep[e.g.][]{Wu79_ECMEtheoryFirstdetail_AKR,1986ApJ...307..808W,Ning21_Z-modeCoales_omgp_by_omgB_less0.3}.}
The ECME emission mechanism predicts that in \nup/\nuB<0.3 X-mode (polarised in the sense of electron gyration), emission is dominant at fundamental \nuB. 
Recent studies using fully kinetic particle-in-cell simulations suggest that a harmonic X-mode emission via Z-mode coalescence can also form at \nup/\nuB<0.3 and be observable~\citep{Ning21_Z-modeCoales_omgp_by_omgB_less0.3}.
Meanwhile, in the 0.3<\nup/\nuB<1.3 range, Z-mode emission at \nuB\ dominates~\citep{Melrose84_ECME_regimes,Winglee85_ECME_regimes_sol_stel_flare} with polarisation in the opposite sense of electron gyration (O-mode). 
However, X-mode emission at higher harmonics of \nuB\ can also form by the coalescence of Z-mode waves when there is a 0.3<\nup/\nuB<1.3 regime~\citep{melrose91_ECME_harmonic_by_coales}.
In a narrow window of 1.3<\nup/\nuB<1.5, X-mode emission emitted directly at the harmonics of \nuB can be observed~\citep{Melrose84_ECMEgrowthrate_damp}.
{X-mode emission at harmonics of \nuB\ are expected via either a direct harmonic emission or various coalescence interactions up to \nup/\nuB<3~\citep{Bastian90_ADLeoflare_arecebo}.} 
Harmonics higher than three are {hardly produced} via any of these process due to the requirement of very high energy density in the unstable wave modes~\citep{melrose91_ECME_harmonic_by_coales}.
So, 3\nuB\ is the highest observable harmonic expected.
Another crucial factor is that the pure X-mode emission at \nuB, or an X-mode emission at harmonics of \nuB, formed by Z-mode coalescence can be nearly 100\% polarised~\citep{Melrose09_CoherentEmiss,melrose2017_coherentEmissMech}.

{When \nup/\nuB > 3, the emissivity of the various ECME processes drops dramatically, and along with the {attenuation due to} gyromagnetic absorption, ECME is expected to be insignificant \citep{Melrose82_ECMEinsun_n_stars,Bastian90_FlarestarsVLA,melrose91_ECME_harmonic_by_coales, Ning21_Z-modeCoales_omgp_by_omgB_less0.3}}.

\noindent {To summarise,} the possible X-mode polarised ECME pathways {driven by loss-cone electron distribution are as follows}:
\begin{enumerate}
    \item {\nup/\nuB<0.3: Fundamental emission at \nuB. Z-wave coalescence could produce emission observable at s\nuB\,(s = 2,3).}
    \item {0.3<\nup/\nuB<1.3: Z or W wave coalescence can produce X-mode emission at s\nuB.}
    \item {1.3<\nup/\nuB<3: Harmonic X-mode emission at s\nuB\ via either a direct emission or wave coalescence interactions.}
\end{enumerate}

\noindent Beyond \nup/\nuB=3, a plasma emission mechanism dominates. \cite{ni20_ECMEinducedplasmaemiss_high_omgp_by_omgB} showed that when \nup/\nuB$>3$, the Z, W, and UH modes triggered by an electron cyclotron maser instability would not escape the plasma. Instead, these waves can interact via a plasma emission mechanism to generate coherent emission {at the fundamental and harmonic of the upper hybrid frequency, $\nu_{\mathrm{UH}}=\sqrt{\nu_{\mathrm{p}}^2 + \nu_{B}^2}$.}

\noindent Besides this, the existing theory suggests that O-mode emission can be produced via the following pathway:
0.3<\nup/\nuB<1.3: Z-mode pathway generates emission at \nuB.

\subsection{The emission mechanism of AD Leo flares}
{In this subsection, we discuss the various possible mechanisms for the observed emission given the possible local physical scenarios during the various activity periods.}
First, we built {1D radial models of density and magnetic field} for the star {and used them to estimate} \nup\ and \nuB\ across coronal heights.
Zeeman-Doppler imaging (ZDI) based studies~\citep{2008MNRAS.390..567M,Bellotti23_ZDIADLeo_Bevol} infer that the magnetic field of AD Leo is predominantly dipolar ($>70\%$), with surface field strengths within $\approx$0.1 - 0.5\,kG. However, {the magnetic field strengths} at strongly flaring active regions can be as high as 3kG~\citep[e.g.][]{Cranmer11_modelMassLoss}. Based on this, we chose a {typical range for field strength at the base of the active region, $\mathrm{B_{foot}}$, between 0.3 and 1\,kG. The chosen range considers a mean ZDI field strength at the lower end and the order of magnitude of the strongest reported active region |B| at the higher end.} 
For the 1D analytical model of the radial magnetic field, two models are considered.
The first one is a dipole-dominated active region magnetic field model suggested for solar active regions by \cite{Asch99_activereg_loopAnalyticalmodel}:
\begin{eqnarray}
   \mathrm{ B(r)=B_{foot}\left(1+\frac{r-1}{0.1}\right)^{-3}}, \label{eqnB1}
\end{eqnarray}
with the dipole scale height, $\mathrm{h_D,}$ assumed to be 0.1\Rs\ (\Rs\ is the stellar radius), based on solar studies~\citep{Asch99_activereg_loopAnalyticalmodel}. We note that r is measured radially from stellar centre in units of \Rs.
In the second model, we assume a simple multipole expansion model of magnetic field with the net magnetic flux assumed to be distributed as 72\% dipole, 21\% quadrupole, and 7\% octapole fraction, based on the ZDI results for observations from 2020~\citep{Bellotti23_ZDIADLeo_Bevol}:
\begin{eqnarray}
    \mathrm{B(r)}= \mathrm{B_{foot}\left(\frac{0.72}{r^3} + \frac{0.21}{r^4} + \frac{0.07}{r^5} \right)}.\end{eqnarray} \label{eqnB2}
{The fractional composition of the magnetic moments employed here are based on large-scale ZDI results that may not be representative of flaring active regions. Since no estimation of multipole moments at the stellar active region exists, we assume the large-scale fractional distribution.
However, the high relative dipolar concentration in Eq.~\ref{eqnB2} assures that this is probably not an uncanny assumption for an active region's magnetic structure if the AD Leo active regions are also similar to those of the Sun with strong bipolar magnetic field concentrations and arcade-like structures.}
The two models are referred to as Model 1 (Eqn.~\ref{eqnB1}) and Model 2 (Eqn.~\ref{eqnB2}) hereafter.

{We derive \nupr\ assuming the density model ($\mathrm{n_e(r)}$) proposed by \cite{2019ApJ...871..214V}:}
\begin{eqnarray}
    \mathrm{n_e(r)} = \mathrm{2.5\times 10^{10} exp(-(r-1)/0.38R_*)\ cm^{-3}}. \label{eqnN}
\end{eqnarray}
{The basal coronal density in the model is derived from OVII line ratios by \citet{Ness04_ADLeodensity}. X-ray observations of pre-flare loops also suggest a basal coronal density within 10$^{10}$ - 10$^{11}\mathrm{cm^{-3}}$~\citep{Namekata20_ADLeoXrayFlares}.} Meanwhile, the assumed radial density profile is empirical, but consistent with the best-fit hydrostatic equilibrium profiles derived for solar active regions based on direct measurements~\citep[e.g.][]{Mann99_Expdensitymodel,Asch01_EUVLoop_density_heating_porofiles}. \cite{Zucca14_densProfile} showed that the data from a {
solar region that hosted a CME}, is modelled well by an exponential hydrostatic equilibrium density profile for r$\lesssim3\mathrm{R_*}$, which is the height range of interest in our study.

\begin{figure}
\centering  
\includegraphics[width=0.4\textwidth]{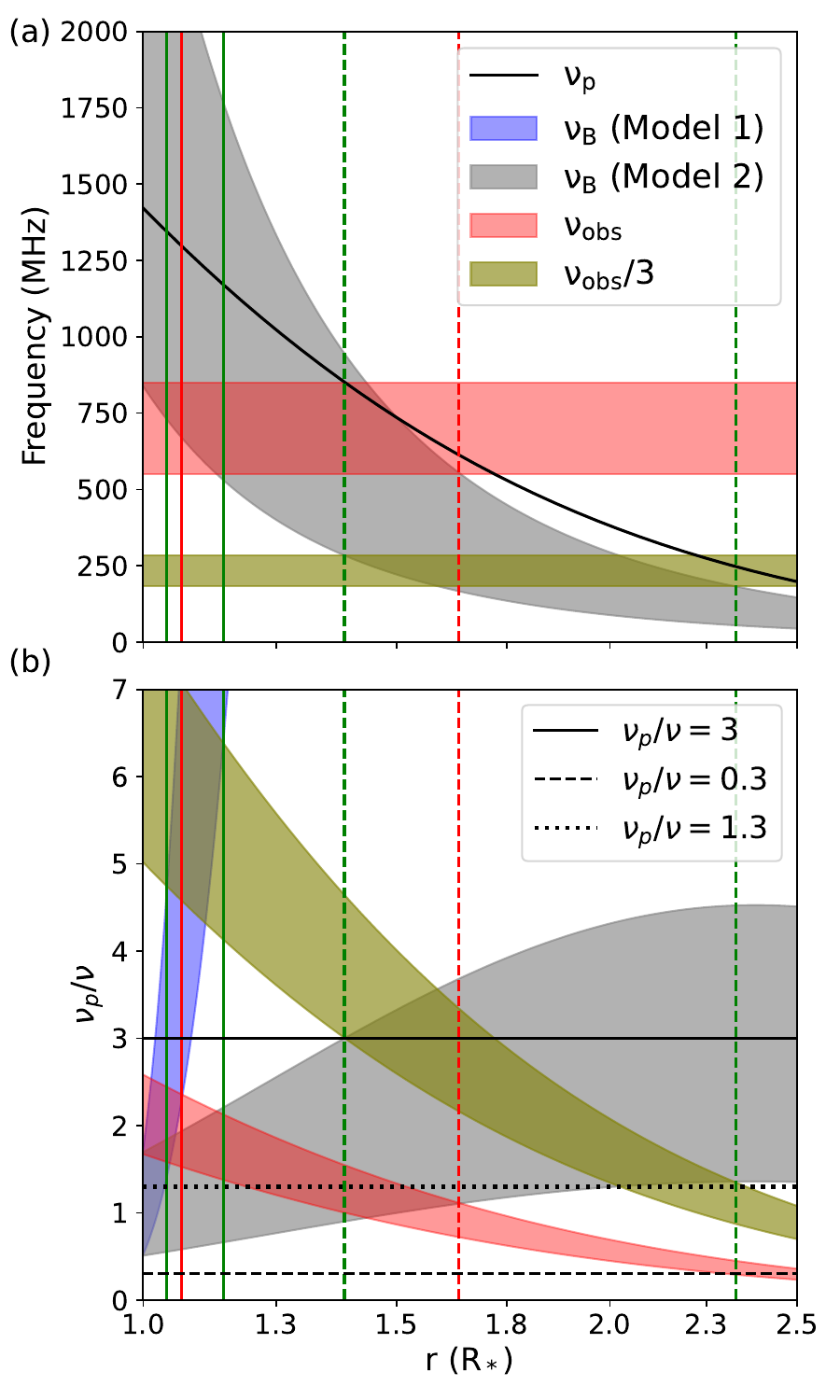}
\caption{Possible scenarios for ECME mechanism to operate. (a): \nupr\ and \nuBr\ for corona of AD\,Leo. \nuB\ is found using Model 1 (Eqn.~\ref{eqnB1}) and Model 2 (Eqn.~\ref{eqnB2}). The observing band ($\mathrm{\nu_{obs}} \in [550,850]$\,MHz) and $\mathrm{\nu_{obs}}$/3 are marked. (b): \nup/\nuBr\ for two B(r) models and the ratios of \nup\ to $\nu$ for \nuobs\ and \nuobs/3 are shown. Colour choice is the same as in the top panel. Red lines mark the r range where $\mathrm{\nu_{obs}}$ overlap \nuB\ and green lines mark the overlap r range of $\mathrm{\nu_{obs}}$/3 with \nuB\,(solid line: Model~1, dashed: Model~2).}
\label{fig5:nup_nub}
\end{figure}

Figure~\ref{fig5:nup_nub}(a) shows the derived \nupr\ and \nuBr. 
The plotted range of \nuBr\ corresponds to 0.3$\leq\mathrm{B_{foot}}\leq$1\,kG. 
The observing band (550 - 850\,MHz) is marked in red. The green band shows the frequencies corresponding to one third of the observation band. 
For an ECME emission process, the observing frequency \nuobs\ = s\nuB,\, where s =1,2,3. {First, we considered Model 1 (blue band)}.
The region where the red band overlaps the blue one corresponds to the r range where \nuobs=\nuB.
Similarly, the r range where the green band overlaps the blue one identifies the r range where \nuobs=3\nuB. Beyond this r range, the possibility of observing ECME in 500 - 850\,MHz is very low since inherent emissivity drops significantly~\citep{Melrose84_ECME_regimes}.
Vertical solid red lines mark the extent of the overlapping r range for \nuobs=\nuB, and the solid green lines mark the r range for \nuobs=3\nuB.
The same exercise is performed for Model 2 {(grey band)}. Instead of solid lines, we used dotted lines of the same colour to mark the r ranges.

\begin{figure}
\centering  
\includegraphics[width=0.4\textwidth]{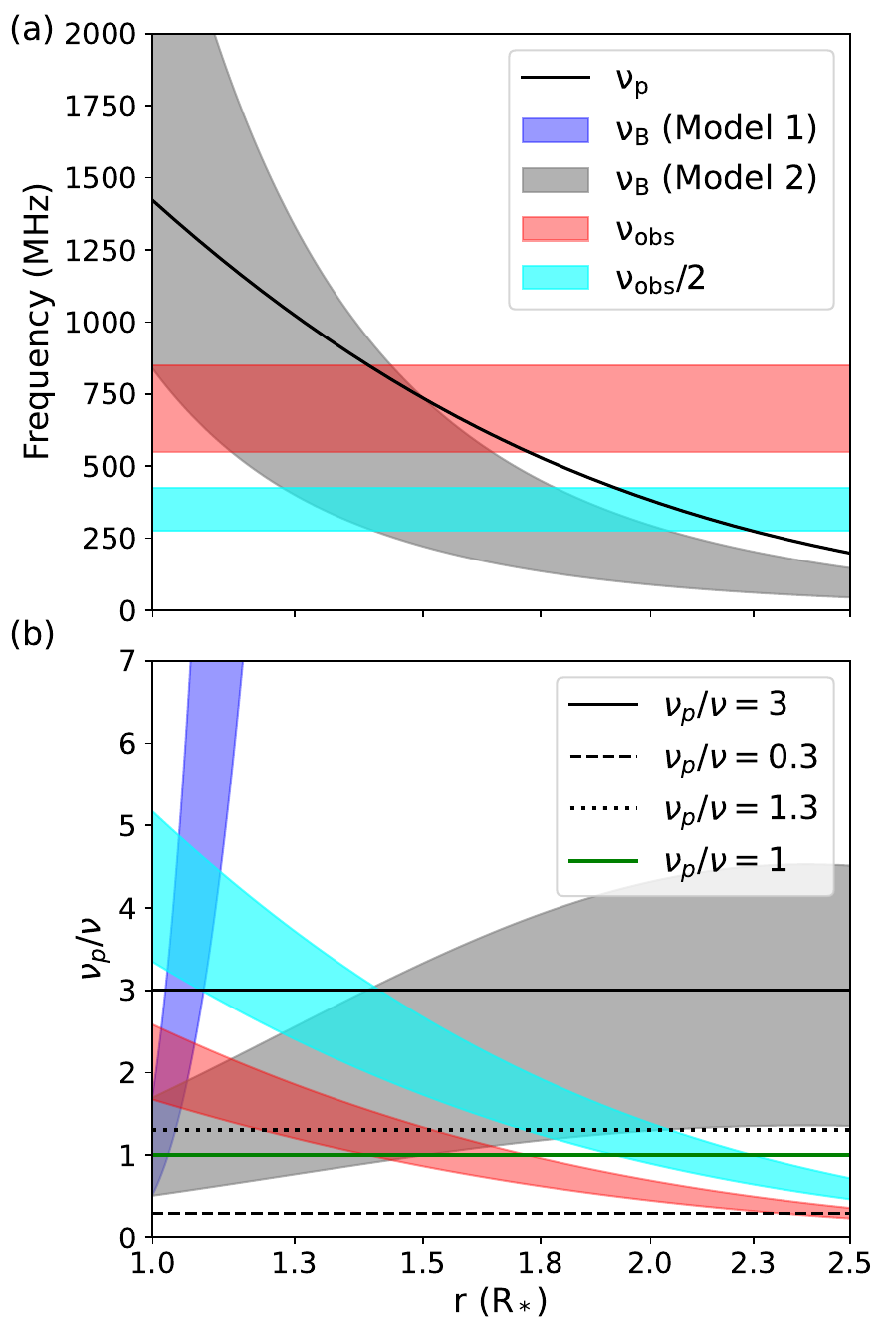}
\caption{Possible scenarios for plasma emission mechanism to operate. (a): Blue, red, and grey bands are the same as in Fig.~\ref{fig5:nup_nub}. Cyan region marks \nuobs/2 band. (b): \nup/\nuBr\ and \nup/\nuobs\ ratio as in Fig.~\ref{fig5:nup_nub}b. Cyan band shows \nup/$\nu$ for \nuobs/2.}
\label{fig6:nup_nu}
\end{figure}

Figure~\ref{fig5:nup_nub}(b) shows the ratio \nup/$\nu$ for $\nu$ = \nuB, \nuobs\ and \nuobs/3. Horizontal lines mark \nup/\nuB=0.3, 1.3, and 3. 
We first consider Model 1. The overlap between the red and blue bands gives us the \nup/\nuB\ values for the r range from where fundamental ECME mode emission can originate in the observing band. 
Clearly this ratio is greater than 1.3, but less than 3, implying that the fundamental X-mode emission is impossible, but a Z-mode ECME is possible.
However, the polarisation will be in the opposite sense of electron gyration (O-mode). 
Moreover, an X-mode polarised emission is possible via the coalescence of Z or W waves, as per the standard theory~\citep{melrose91_ECME_harmonic_by_coales}. Yet, the resultant harmonic X-mode emission would not be observable in our \nuobs\ band, but in the 2\nuobs\ band.
{Meanwhile, the overlap between green and blue bands gives us the \nup/\nuB\ values in the r range from where, if third harmonic emission arises, it can be observed in our \nuobs.}
The range of \nup/\nuB\ is much greater than~three, making an ECME emission mechanism unfavourable. 
Similar to the above discussion, for Model 2, the intersection between the grey and red bands gives the range of \nup/\nuB\ for fundamental ECME emission, and the overlap between the grey and green bands gives the range of the ratio for harmonic emission to be observable in \nuobs\ band. 
If we consider the grey and red band overlap, we have cases for 0.3<\nup/\nuB<3. Either an O or Z mode emission (0.3<\nup/\nuB<1.3) or a harmonic X-mode (1.3<\nup/\nuB<3) can be triggered here. However, akin to the case of Model 1, the harmonic X-mode emission born out of Z- or W-mode coalescence or a direct emission cannot be observed in our observing band. 
However, the overlap between the green ($\nu$= \nuobs/3) and grey bands falls within 1.3<$\nu$/\nuB<3, favouring a harmonic X-mode ECME from its r range via Z- or W-mode coalescence or a direct harmonic emission, which is observable in the \nuobs\ band.

{To summarise ,  Model 1 only favours an O-mode polarised (opposite sense of electron gyration) emission at \nuB\ via a Z-mode pathway, which is observable in the \nuobs\ band. However, Model 2 favours an X-mode (polarised in the sense of electron gyration) emission at harmonics of local \nuB\ formed via either a direct emission or Z/W-mode coalescence pathways and observable in the \nuobs\ band.}

We now consider the polarisation in the bursts to further constrain the mechanism. The polarisation has consistently been left circular with Stokes V/I > 60\% during bursts. At higher frequencies ($\gtrsim$700\,MHz), we find higher polarisation degrees attained during strong flares (e.g. Fig.~\ref{fig3:V_spectra}).
\cite{2008MNRAS.390..567M} had shown that the south pole of AD Leo is inclined well along the line of sight (LOS) at an angle of 20$^\circ$. So, there is a high chance that magnetic field structures could be oriented southwards on the LOS. This means the observed emission is in the direction of electron gyration,which means the emission is in X-mode.
Also, the degree of polarisation is $>60\%$ during flares, often reaching $\approx$ 100\% at higher frequencies (see  Fig.~\ref{fig3:V_spectra}).
{If the emission is via an ECME process, the possible pathways are  either a fundamental X-mode ECME mechanism from regions where \nuobs=\nuB\ or a harmonic X-mode emission from regions where \nuB=\nuobs/s and 0.3<\nup/\nuB<3 (s=2,3)}.
Based on the inferences from Fig.~\ref{fig5:nup_nub}, Model 1 does not favour both fundamental and harmonic X-mode ECME, but Model 2 favours a harmonic X-mode emission by all possible pathways. {Meanwhile, ~\cite{Jiale2023_FAST_ADLeoflares} observed the star in the 1 - 1.5\, GHz band during the strong activity period, from 21:13 – 22:48 UT, and reported emission features that are better explained by an ECME mechanism. Unfortunately, their observations did not extend further into the weak and quiescent activity phases.}

{{Our analysis did not} explicitly consider the propagation effects that {would have affected} the observed ECME. Without multi-waveband spectroscopic data, the existing stellar atmospheric models~\citep[e.g.][]{Fontenla16_GJ832_dM1Datmos,Sven13_ADLeoChromModel} cannot be well constrained to derive a reliable ambient/quiescent physical structure (especially B(r) and $\mathrm{n_e(r)}$), which is crucial to assessing the atmospheric opacity~\cite[see][]{white97_gyroresonance_emiss_Rev,Treumann06_ECME_in_astro_Rev}. Consequently, based on current  knowledge on the ECME mechanism as derived from solar observations, the emissivity should fall drastically above $s=2$  (see Sect.~\ref{sec:ECME}). Moreover, the emission for higher harmonics (s=$2,3$) is less affected by gyromagnetic absorption from upper coronal layers~\citep{Zheleznyakov70_radioemiss_mech_book,Melrose82_ECMEinsun_n_stars}. Both of these strongly favour the detection of ECME at s=2,3.
Our inferences on ECME generation based on 1D stellar models also favour harmonic emission.}

Plasma emission, which is the competing mechanism to ECME, produces coherent emission observable at \nup\,(F-branch) and 2\nup\,(H-branch)~\citep{ginzburg1958,Tsytovich69, melrose1972,Melrose09_CoherentEmiss}. This mechanism generally favours polarisation in the O-mode~\citep{melrose2017_coherentEmissMech} especially for the F branch.
For the H branch, this is true if \nup$>$\nuB, but if \nup/\nuB$\lesssim$1.5, highly polarised X-mode emission at \nuobs=2\nup\ is possible~\citep{Willes97_2ndHarm_poln_strng_weakBfield}. 
Given that we observe a high \TB\ X-mode polarised flares, it is likely a H-branch emission from regions with \nup/\nuB$\lesssim$1.5.
{From Fig.~\ref{fig6:nup_nu}b, we infer that the H-band emission observable in \nuobs\ should arise from r$\gtrsim$1.9\Rs, where \nup<\nuobs/2.} In this r range, \nup>>\nuB\ for Model 1. However, for some B$_\mathrm{foot}$ values, \nup/\nuB<1.5 is satisfied for Model 2. So, only Model 2 can support an H-band plasma emission in X mode.

{Based on the discussions so far, Model 2 favours X-mode polarised emission via both H-branch plasma emission mechanisms and ECME emission processes at harmonics of \nuB.} 
We note that Model 1 does not support any of the likely emission processes favouring a strong X-mode ECME or H-band plasma emission. This possibly means that a solar-like active region model is not good for M~dwarfs, but a ZDI-based multipole expansion model works better.
For the rest of the discussion, we hence only consider Model 2.
Meanwhile, the H-branch plasma emission observed in the Sun has always shown polarisation levels below 30\%~\citep{Willes97_2ndHarm_poln_strng_weakBfield,Hilaire,Rahman20_typeIIIstats}.
So, during a strong activity period, given the high polarisation of 60 - 100\% (Figs.~\ref{fig1:DS_LC}-\ref{fig3:V_spectra}), the emission mechanism is likely via an ECME mechanism.

{In the analysis so far, we had kept the basal coronal density constant in the model. We now consider a variation of this value within an order of magnitude and assess the possible ECME pathways supported by the two B(r) models. {For this analysis, we vary n$_\mathrm{e}$ within 10$^{10}$ - 10$^{11}$\,cm$^{-3}$, consistent with the values reported by previous spectroscopic and X-ray continuum observations. This corresponds to a variation in the assumed basal coronal density in Eq.~(\ref{eqnN}) by 0.4 to 4 times.} If the basal density goes up four times, since \nup$\approx \sqrt{n_e}$ we would see an increase in \nup\ by a factor of two, which in turn raises all the \nup/$\nu$ bands in Fig.~\ref{fig5:nup_nub} by the same factor.
We note that in Fig.~\ref{fig5:nup_nub} the green band overlaps with the grey one (Model 2) within 1<\nup/$\nu$<3.
So, the harmonic X-mode ECME emission pathway, previously supported by Model 2, will remain feasible, but now in the 2<\nup/$\nu$<3.
{The other ECME pathways that were already not supported by the previously assumed lower density plasma will be unsupported by a denser plasma due to enhanced damping of the ECME driving instability.}
However, since \nuobs/3 overlaps with Model 2 in \nup/$\nu$<10, an ECME-driven coherent plasma emission process suggested by ~\cite{ni20_ECMEinducedplasmaemiss_high_omgp_by_omgB} can produce harmonic X-mode emission observable in \nuobs.
{Meanwhile, if we assume a 0.4 times lower density in the active region, all the \nup/$\nu$ bands in Fig.~\ref{fig5:nup_nub} will be reduced by a factor of 1.6.} In this case for Model 2, harmonic X-mode emission will be observable as before. {Interestingly, a portion of the overlapping region between Model 1 (blue band) {and} the green band will fall within 2.5<\nup/$\nu$<3, suggesting a possibility of observing a harmonic X-mode emission in \nuobs. This may make Model 1 viable for a small range of B$_{foot}$ and basal density values.} However, Model 2 consistently supports an ECME mechanism for an order of magnitude range of B$_{\mathrm{foot}}$ and n$_e$ values.}


\subsection{Weak activity period {and the }type-IV burst}
\begin{figure*}
\centering  
\includegraphics[width=0.65\textwidth,height=0.62\textheight]{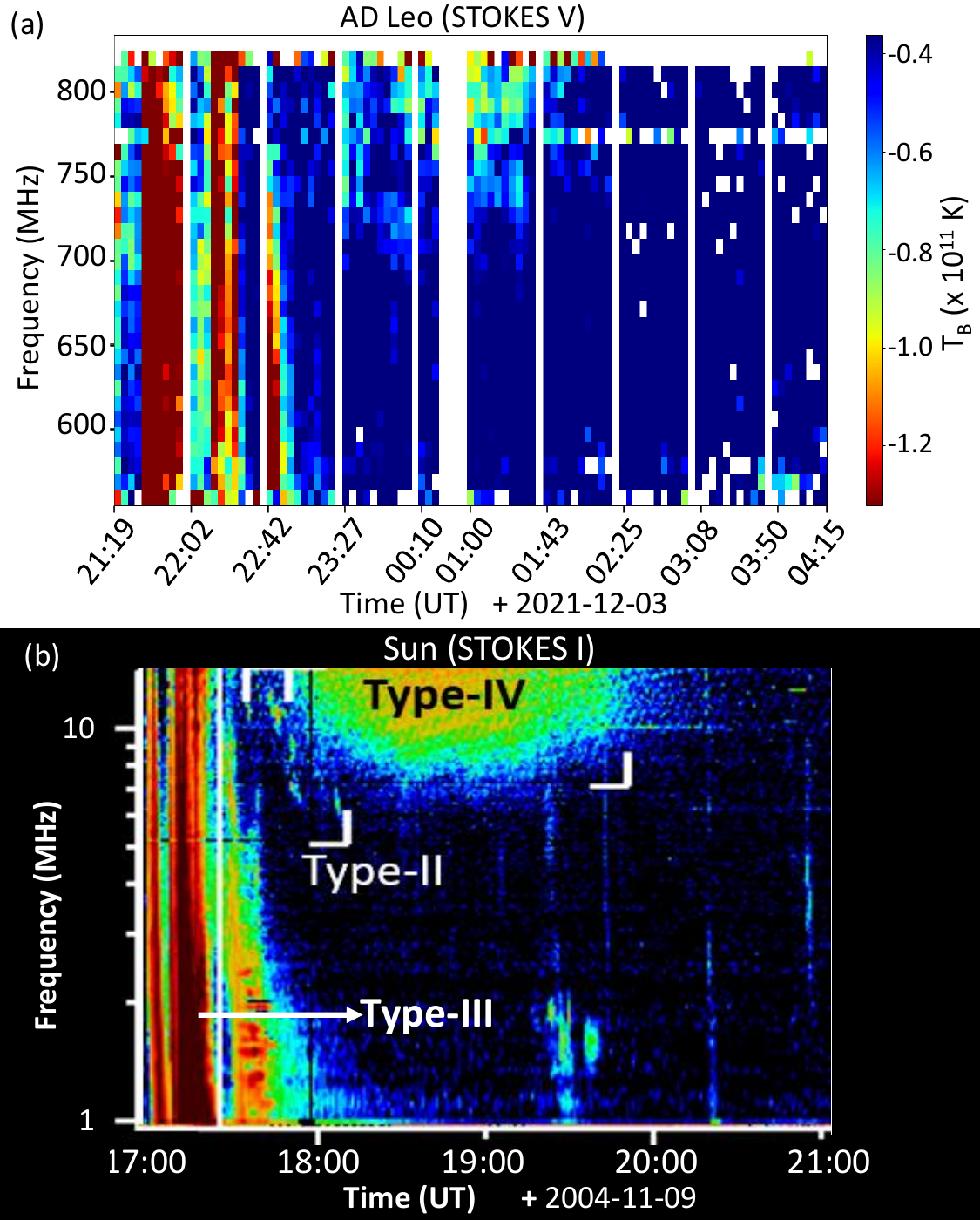}
\caption{Comparison between AD Leo DS and a typical solar DS during a CME. (a): AD Leo VISAD spectrum with 200\,s and 10\,MHz averaging with a saturated colour scale. (b): A solar radio spectrum from Wind/WAVES instrument showing the different burst types associated with a solar CME on Jan 4, 2004. Solar type-IV and type-III features can be identified in the AD Leo spectrum. Solar data credit: CDAW data centre.}
\label{fig7:VISAD_avg}
\end{figure*}
The post-flare weak activity phase is found to last for $\approx$\,1.4\,h.
The Stokes V dynamic spectrum (Fig.~\ref{fig1:DS_LC}a) shows that the post-flare emission {shows a bright patch confined to frequencies $\gtrsim$ 700\,MHz, between 2021-12-03 23:27 and 2021-12-04 02:25 UT. To highlight this better, a 200\,s averaged VISAD spectrum was made (see Fig.~\ref{fig7:VISAD_avg}a).
The imaging-based spectrum from the weak activity period (Fig.~\ref{fig3:V_spectra}b) also shows a characteristic high-polarisation bright-emission feature above $\approx$ 700 - 730\,MHz, where the emission polarisation jumps above $\approx$ 55\% and rises to $\approx$ 80\% as frequency increases.
As per the solar radio burst classification scheme, the aforementioned features suggest that the weak activity phase corresponds to a type-IV radio burst confined to frequencies above 700\,MHz~\citep{Gopal11_PREconf,Hillaris16_typeIV,Salas20_typeIVPoln_struct}.
For comparison, Fig.~\ref{fig7:VISAD_avg}b shows a solar dynamic spectrum during a strong CME taken by the Wind/WAVES instrument. The solar data are taken from the publicly available type-IV catalogue compiled by Coordinated Data Analysis Workshops (CDAW) data centre~\footnote{\MYhref{https://cdaw.gsfc.nasa.gov/CME\_list/radio/type4/DHtypeIV_catalog.html}{https://cdaw.gsfc.nasa.gov/CME\_list/radio/type4}}.
The solar event was so chosen to match the dynamic spectral emission profile of the AD Leo. The morphological analogues of the type-III and type-IV bursts in the solar template spectrum can be identified in the AD Leo DS.
}

Type-IV radio bursts are associated with strong flares and CMEs in the Sun, as the sought-after type-II bursts~\citep{Robinson86_typeIV_fastCME_typeII_link,Gopal11_PREconf,Hillaris16_typeIV,Anshu21_CME_typeIV}.
While a type-II burst is associated with the coronal shock generated by a fast CME, a type-IV burst is generated by flare-accelerated electrons trapped either in the post-flare loops or in the moving plasmoids associated with the CME~\citep[e.g.][]{wild1970,Smerd70_curvingtypeII_imaging,Gopal05_typeII-m-DH_SEPlink,white2007_radiobursts,Gopal11_PREconf}.
Type-IVs are generally associated with relatively longer and stronger solar X-ray flares than type-IIs~\citep{Cane88_SXR_metricburst_props}.
Also, type-IV bursts are often linked to strong CMEs, which expand out to heliocentric angular widths $\gtrsim$60$^{\circ}$, with mean speeds $\gtrsim$900\,km\,s$^{-1}$~\citep{Robinson86_typeIV_fastCME_typeII_link,Gopal11_PREconf,Hillaris16_typeIV}. 
Recently, \cite{Zic20_typeIV_ProximaCen} reported the first type-IV burst detection in a star other than the Sun, namely Proxima Centauri, which is an old ($\approx$5\,Gyr) M~dwarf.
Given its age ($\approx$ 250\,Myr) and rotation period (P$_\mathrm{rot}$ = 2.2\,days), AD\,Leo belongs to a fast rotating (period $<$ 5\,days), young (age $<$1\,Gyr) and more active stellar population (`C' branch) compared to the the Sun (age = 4.6\,Gyr; P$_{\mathrm{rot}}$ = 25\,days), which belongs to the less active `I' branch stars~\citep{Barnes03_Rot_Vs_age_Vs_Activity,2014ApJ...789..101B}.
\cite{Barnes03_Rot_Vs_age_Vs_Activity} demonstrated that the stars undergo an evolutionary migration from `C' to `I' branch as they spin down with age due to mass and angular momentum loss. This migration involves a fundamental change in the way the internal dynamo couples with the external active atmospheric layers and the stellar wind~\citep{2014ApJ...789..101B,2018ApJ...862...90G}.
So, though Proxima Centauri is an M~dwarf, it is an `I' branch star, but AD\,Leo is not.
Here, we report the first detection of a type-IV burst on a young ($\approx$250\,Myr), active `C' branch M~dwarf.

We do not find evidence of a type-II burst, making it still elusive in an M~dwarf despite monitoring campaigns spanning several days on various stars, including AD\,Leo~\citep[e.g.][]{Osten2008,2019ApJ...871..214V}.
Though rare, solar type-II bursts have also been found to be associated with coronal shocks produced by failed eruptions, unrelated to CMEs~\citep[e.g.][]{Magdaleni12_typeII_noCME,Morosan23_typeII_noCME}.   
AD Leo is an active star with a magnetic field that is an order of magnitude stronger and predominantly axi-symmetric~\citep[][]{2019ApJ...877..105M}. Theoretical models have argued that the strong magnetic field structures and the associated high Alfv\'{e}n speed barriers in the stellar corona could either block even the energetic flares from causing a CME (failed eruption scenario), or resist the formation of a strong shock required to initiate a type-II emission during an eruption/CME~\citep[][]{2020MNRAS.494.3766O,2022AN....34310100A}. 
This means that, unlike the case of the Sun, AD Leo could host more failed eruptions during strong flares, and even if a CME or eruption occurs, type-II bursts need not be triggered for most events.
These scenarios can explain why there have been no reports of a type-II burst on AD\,Leo despite several monitoring campaigns~\citep[e.g.][]{Osten2006,Osten2008,2019ApJ...871..214V}.
In any case, since post-flare loop structures will trap flare-accelerated electrons, strong type-IV bursts could trace CMEs better than type-IIs in M~dwarfs.

Given the high X-mode polarisation level and high \TB\ at \nuobs$\gtrsim$700\,MHz, and based on our inferences for the active region in the previous section, the type-IV burst emission likely originates from an ECME mechanism via either a Z/W-mode coalescence or a harmonic emission pathway. 
An ECME mechanism is usually triggered by an unstable loss-cone distribution of energetic electrons formed within a magnetic bottle-like configuration, readily provided by the post-flare loops~\citep[e.g.][]{Carley19_ECMZmode_loss-cone}.
In the case of the Sun, type-IV bursts are generally associated with plasma emission mechanism in the metric waveband~\citep{Weiss63_typeIV_plasmaEmiss,Salas20_typeIVPoln_struct}.
However, during some strong flares, associated type-IV bursts have demonstrated $\approx$ {millisecond to second} scale spiky emission features in their dynamic spectra with high degrees of circular polarisation and \TB, often modelled as generated by an ECME mechanism~\citep[e.g.][]{Carley19_ECMZmode_loss-cone,Liu18_ECM_typIVstat,Morosan19_typeIV_emissmech,2022LRSP...19....2C}.
Certain high time-resolution studies of AD Leo in 1-3 GHz observations have reported $\approx$ms scale spiky emission features~\citep[e.g.][]{Osten2008,Jiale2023_FAST_ADLeoflares}.
{The time resolution of uGMRT data} restricts us from exploring sub-second-scale fine structures in the dynamic spectrum. 
The ECME signatures we detect in the type-IV emission above 700\,MHz could probably be formed by numerous sub-second-scale fine structures. 
Meanwhile, at frequencies below 700\,MHz the emission polarisation is well below 30\%.
The emission mechanism here is likely a H-branch plasma emission, which is known to produce high \TB\ emission at low polarisation levels~\citep[e.g.][]{Willes97_2ndHarm_poln_strng_weakBfield,Hilaire,Rahman19_corhole_emiss}.

Since the type-IV emission is seen to be mostly confined within 700\,MHz (Fig.~\ref{fig7:VISAD_avg}), using Model 2 and {assuming that the ECME is formed at either the second or the third harmonic of \nuB, the height of the post flare loop can be inferred. 
It can be seen from Fig.~\ref{fig6:nup_nu}b that within 1.8 - 2.25\,\Rs, the overlap region of the cyan and grey bands satisfy the condition of 0.3<\nup/\nuB<1.5 required for Z-mode or harmonic X-mode mechanism to operate efficiently and produce a second harmonic ECME observable in \nuobs.}
Since 700\,MHz is roughly the centre of the cyan band, the post-flare loop height could be $\approx$ 1.8 - 2.0\,\Rs. 
Similarly, if the emission is at the third harmonic (\nuobs = 3\nuB), we should use Fig.~\ref{fig5:nup_nub}(a-b) to obtain emission height. Here, instead of the cyan band, we considered the overlap between the green and the grey bands. The emission height range turns out to be $\approx$ 2.0 -2.5\,\Rs, implying a post-flare loop height of $\approx$2 - 2.25\,\Rs.
This implies that the star can have loops rising to about 1\Rs\ above the photosphere.


\subsection{Quasi-steady period}
The quiescent {Stokes I flux level varies within $\approx$ 1 - 1.75\,mJy (see  Figs.~\ref{fig1:DS_LC}c \& ~\ref{fig2:LC_imging}), implying \TB\ values of the order of $\approx$ 10$^{10}$\,K.}
However, the polarisation levels of $\approx$20 - 30\% inferred from the light curve suggests a H-band coherent plasma emission mechanism.
{The observed emission is likely a superposition} of various spatially separated coherent plasma emission regions or a mixture of gyrosynchroton and plasma emission regions  ~\citep{Bastian90_ADLeoflare_arecebo,Bastian94_stellarflaresRev}.


\section{Conclusions}
\label{sec:conclusion} 

{We present a two-hour-long metric band flaring activity in a young active M~dwarf, AD Leo. {A hydrostatic equilibrium electron density profile was assumed with a basal density ranging within 10$^{10}$ - 10$^{11}$\,cm$^{-3}$}. Based on the existing knowledge on the stellar magnetic field structure, different radial field profiles (B(r)) were considered: a solar-like model and a simple multipole expansion model. Contrasting the radio data with these models, we conclude the following:}
\begin{itemize}
    \item {The active region B(r) is more consistent with a multipole expansion model than a solar-like model, given the possible range of $|$B(r)$|$ and basal coronal density.}
    \item {The observed radio bursts are powered by an ECME mechanism.}
\end{itemize}

The period of strong flaring was succeeded by a post-flare activity period. We report a solar-like type-IV radio burst in AD Leo, in the post-flare phase, {characterised by a distinctive dynamic spectral morphology and circular polarisation levels ranging from 60\%\ - 100\%.} Previously, a type-IV emission was reported in Proxima Centauri by \cite{Zic20_typeIV_ProximaCen}.
This is the first solar-like type-IV detection on a young active M~dwarf belonging to a different age-activity population (`C' branch) relative to the Sun (belonging to `I' branch)~\citep{Barnes03_Rot_Vs_age_Vs_Activity}.
The emission characteristics favour an ECME mechanism {at the second or third harmonic of the local gyrofrequency.}
The detection of the type-IV burst could signify a strong CME, since solar type-IVs are known to be associated with fast and wide CMEs. Nevertheless, an hour-long type-IV burst signifies powerful accelerated particle activity in the post-flare loops.
Based on the frequency extent of the type-IV burst in the VISAD dynamic spectrum, we infer that the post-flare loops could rise up to $\approx$1\,\Rs\ scales above photosphere into the inter-planetary space.
The quasi-steady emission that followed the type-IV activity is likely powered by a mixture of several coherent plasma emission regions.

\section*{Acknowledgements}
This work is supported by the Research Council of Norway through the EMISSA project (project number 286853) and the Centres of Excellence scheme, project number 262622 (``Rosseland Centre for Solar Physics''). We thank the
staff of the GMRT that made these observations possible. GMRT is run by the National Centre for Radio
Astrophysics of the Tata Institute of Fundamental Research. This research made use of NASA's Astrophysics Data System (ADS). 
AM and NG are supported in part by NASA's STEREO project and LWS program.
AM acknowledges the developers of the various Python modules namely Numpy \citep{numpy}, Astropy \citep{astropy}, Matplotlib \citep{matplotlib} and multiprocessing. AM also thanks the developers of CASA \citep{casa}.
AM acknowledges the discussions with Barnali Das which greatly helped speed up the data analysis. AM acknowledges the discussions with Tim Bastian.
\bibliographystyle{aa}
\bibliography{EMISSA_allref}


\end{document}